\documentclass[prl,floatfix,twocolumn]{revtex4-1}

\usepackage[english]{babel}
\usepackage{amsmath,amsfonts}
\usepackage{wasysym}
\usepackage{bm}
\usepackage{graphicx}
\usepackage{dcolumn}
\usepackage{color}
\usepackage{nicefrac}
\usepackage{ifpdf}

\DeclareMathOperator{\tr}{tr}

\def\dashedrule#1#2#3{{%
\dimen1=#2 \divide\dimen1 by 2
\def\@ruledash{%
  \rule{\dimen1}{0pt}%
  \rule[0.5ex]{#1}{0.4pt}%
  \rule{\dimen1}{0pt}}%
\count1=0
\loop%
\ifnum\count1<#3%
\advance\count1 by 1%
\@ruledash%
\repeat}}

\begin{document}

\title{Geometric phase effects in dynamics near conical intersections:
  Symmetry breaking and spatial localization}

\author{Ilya G. Ryabinkin} 
\affiliation{Department of Physical and Environmental Sciences,
  University of Toronto Scarborough, Toronto, Ontario, M1C 1A4,
  Canada; and Chemical Physics Theory Group, Department of Chemistry, University of Toronto, Toronto,
  Ontario, M5S 3H6, Canada}

\author{Artur F. Izmaylov} 
\affiliation{Department of Physical and Environmental Sciences,
  University of Toronto Scarborough, Toronto, Ontario, M1C 1A4,
  Canada; and Chemical Physics Theory Group, Department of Chemistry, University of Toronto, Toronto,
  Ontario, M5S 3H6, Canada}

\date{\today}

\begin{abstract}
We show that finite systems with conical intersections can exhibit spontaneous symmetry breaking which
manifests itself in spatial localization of eigenstates. This localization has 
a geometric phase origin and is robust against variation of
model parameters. The transition between
localized and delocalized eigenstate regimes resembles a continuous
phase transition. The localization slows down the low-energy quantum nuclear
dynamics at low temperatures.
\end{abstract}


\maketitle

A conical intersection (CI) of several electronic states is one of the
most common structural motifs in molecules where the
Born--Oppenheimer (BO) approximation breaks down~\cite{Migani:2004/271,
  Yarkony:1996/rmp/985}. Due to energetic proximity of potential
energy surfaces, nuclear motion near CI triggers electronic transitions. 
These transitions are not the only effect that CIs
produce: parametric evolution of the eigenstates of the
\emph{electronic} Hamiltonian along a closed path encircling a locus of CI
gives rise to an extra $(-1)$ phase factor accumulated by the
intersecting electronic eigenstates~\cite{LonguetHigg:1958/rspa/1,
  Berry:1984/rspa/45}. This additional phase, termed the
\emph{geometric phase} (GP)~\cite{Mead:1979/jcp/2284,
  Berry:1984/rspa/45, Berry:1987/rspa/31} does not depend on a size
or a shape of the encircling loop, provided that no other degeneracies
are enclosed. GP also affects the \emph{nuclear} motion, because
changing the sign of the electronic wavefunction will necessarily
change the sign of the nuclear wavefunction in order for the \emph{total}
wavefunction to be single-valued. GP effects were found to be
crucial for modeling vibrational spectra of Jahn-Teller distorted
compounds (e.g., Na$_3$)~\cite{Kendrick:1997/prl/2431,
  Ham:1987/prl/725, Child:2003} and cross sections in 
  low-energy atom-molecule reactive scattering (e.g., $\rm H +
O_2$)~\cite{Kendrick:1996/jcp/7475, Kendrick:1996/jcp/7502,
  Kendrick:2000/jcp/5679, Kendrick:2003/jpca/6739,
  JuanesMarcos:2005/sci/1227}.

In this Letter, we report yet another remarkable GP effect: 
spontaneous symmetry breaking that manifests itself in spatial
localization of low-energy eigenstates. It is common in quantum mechanics
that low-lying eigenstates are delocalized over all
energetically-accessible regions. Delocalization lowers the kinetic
energy of a system and, if it is not counteracted by the potential,
lowers the total energy. However, in the presence of GP some
eigenstates are found to be immune to delocalization even though
the potential does not counteract.  Although this localization can be
seen as a consequence of destructive interference between
different tunneling paths connecting energetically-accessible regions
in systems with CI~\cite{Schon:1995bi, Garg:1993/epl/205,
  Garg:2000/epl/382, Kececioglu:2001/prb/064422, Stelitano:1995tj}, we
show that the destructive interference alone is not
sufficient for the localization.

We consider a generic two-state (``full'') model exhibiting CI along
with its two single-surface approximations: BO and BO+GP. The BO model
neglects GP completely, whereas the BO+GP model uses a
position-dependent phase factor $e^{i\theta(\mathbf{x})}$ that
changes the sign of a nuclear wavefunction upon encircling 
CI~\cite{Mead:1979/jcp/2284, Mead:1992/rmp/51}. The two-state model
includes both GP and non-adiabatic transitions, therefore comparing results 
from all three models allows us to isolate and quantify pure GP effects.

As we show below, the spinor symmetry of the full model is
preserved by the BO+GP model but is lost in the BO approximation. The spinor
symmetry gives rise to \emph{degeneracy} of the ground state for some
values of model parameters and thus produces the quantum phase transition~\cite{Sachdev2011:Quant_phase_trans}. Small variations of parameters
can lift the degeneracy, but associated localization does not
disappear. Further parameter variation eventually 
destroys the localization, and critical values for the parameters  
form a phase diagram. The phase diagram qualitatively explains
observed differences in nuclear dynamics at low energies with and
without GP.

\begin{figure}
  \centering
  \includegraphics[width=0.5\textwidth]{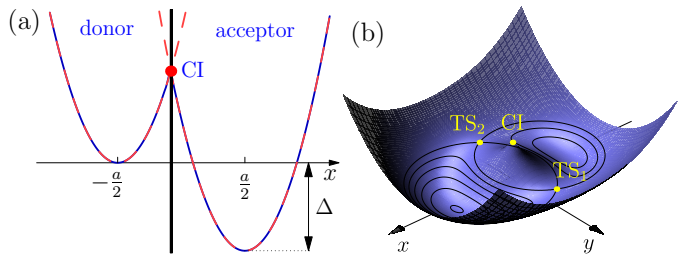}
  \caption{ a) Diabatic potentials (red dashed lines) $V_{11}$ and
    $V_{22}$, and the lowest adiabatic potential $W_{-}$ (blue solid line), in $y=0$ section. 
    The thick black line separates donor and acceptor wells. b) The lowest adiabatic potential $W_{-}$ with two symmetric transition states 
    labelled as TS$_1$ and TS$_2$.}
  \label{fig:2d-sketch} 
\end{figure}

\textit{Model.---} In a diabatic representation, the Hamiltonian of the
two-state CI model is
\begin{equation}
  \label{eq:diab-2st}
  {\hat H} = {\hat T}_N \mathbf{1}_2 + 
  \begin{pmatrix}
    V_{11} & V_{12} \\
    V_{12} & V_{22}
  \end{pmatrix},
\end{equation}
where ${\hat T}_N$ is the nuclear kinetic energy operator,  
$V_{11}$ and $V_{22}$ are diabatic potentials
represented by identical 2D parabolas shifted in space  
 and in energy (Fig.~\ref{fig:2d-sketch}a) and coupled by the 
 $V_{12}$ potential ($\hbar=1$)
\begin{align}
  \label{eq:diab-me-11}
  V_{11} = {} & \dfrac{\omega_1^2 }{2}\left(x + \dfrac{a}{2}\right)^2
  + \dfrac{\omega_2^2
  }{2}y^2 + \Delta/2, \quad  V_{12} = c y, \\
  \label{eq:diab-me-22}
  V_{22} = {} & \dfrac{\omega_1^2}{2} \left(x - \dfrac{a}{2}\right)^2
  + \dfrac{\omega_2^2}{2} y^2 - \Delta/2.
\end{align}
In spite of its simplicity, $N$-dimensional generalization of this 2D model 
was successfully applied to modeling vibronic spectra of real molecules 
with CIs \cite{Koppel:1984/ACP/59,Worth:2004/ARPC/127}.
Diagonalization of the two-state potential matrix in
Eq.~(\ref{eq:diab-2st}) gives the following adiabatic potentials
\begin{align}
  \label{eq:Wmin}
  W_{\pm} = & {} \dfrac{1}{2}\left(V_{11} + V_{22}\right) \pm
  \dfrac{1}{2}\sqrt{\left(V_{11} - V_{22}\right)^2 + 4 V_{12}^2}.
\end{align}
We focus on a lower-surface Hamiltonian ${\hat H}_{-} = {\hat P}^2/{2}
+ W_{-}$ (Figs.~\ref{fig:2d-sketch}a-b), which governs the dynamics in
the BO and BO+GP approximations. The BO and BO+GP models differ in form of
${\hat P}$: ${\hat P}^\text{BO} = -i\nabla$ for the BO model
[$\hat{H}_{-}^{\rm BO}=\hat{H}_{-}({\hat P}^\text{BO})$], and 
${\hat P}^\text{GP} = -i e^{-i\theta} \nabla e^{i\theta}
\equiv -i\nabla + \nabla\theta$ for the BO+GP model
[$\hat{H}_{-}^{\rm GP}=\hat{H}_{-}({\hat P}^\text{GP})$]. The latter ${\hat P}^\text{GP}$ identity
shows that one can still work with \emph{single-valued} wavefunctions
even in the presence of GP at expense of an extra term in definition
of the canonical momentum~\cite{Wittig:2012/pccp/6409}.
$\theta=\theta(x, y)$ is defined as the mixing angle of the two-state unitary
transformation that diagonalizes the potential matrix
in Eq.~(\ref{eq:diab-2st})~\footnote{Care must be taken when
  arguments of $\arctan(y/x)$ cross the coordinate axes. A convenient
  choice is to allow the functional discontinuities from $\pi$ to $-\pi$ upon
  crossing the negative $x$ axis in the counterclockwise direction.}
\begin{equation}
  \label{eq:theta-explicit}
  \theta(x,y) = \frac{1}{2}\arctan \dfrac{2\,V_{12}}{V_{11} -
    V_{22}} = \frac{1}{2}\arctan \dfrac{\gamma y}{x - b}, 
\end{equation}
where $b = \Delta/(\omega_1^2 a)$ is the $x$-coordinate of the CI point, and $\gamma
= {2c}/{\omega_1^2a}$ is dimensionless coupling strength.

\textit{Symmetry and spectrum degeneracies.---} For $\Delta = 0$ and
arbitrary values of other parameters, the potential $W_{-}$ possesses
$C_{2v}$ symmetry, which is also the symmetry of the BO
Hamiltonian $\hat{H}_{-}^{\rm BO}$. In contrast, $\hat{H}_{-}^{\rm GP}$ and $\hat{H}$ have
the \emph{double} group symmetry $C_{2v}^\dagger$~\cite{Altmann2005:Rotations} 
that adds to the $C_{2v}$ elements an extra rotation by $2\pi$ ($R$).  
$R$ acts nontrivially only on
the double-valued eigenfunctions of $\hat{H}_{-}^{\rm GP}$ and
two-component eigenfunctions of $\hat{H}$.  $C_{2v}^\dagger$ is a
non-Abelian group, and all eigenfunctions of $\hat{H}_{-}^\text{GP}$
and $\hat{H}$ transform according to its two-dimensional irreducible
\emph{spinor} representation $E_{1/2}$, giving rise to a
doubly-degenerate spectrum. By allowing $\Delta \ne 0$, we lower the
symmetry to $C_s$ for $\hat{H}_{-}^{\rm BO}$ and to $C_{s}^\dagger$ for
$\hat{H}_{-}^{\rm GP}$ and $\hat{H}$.  Both $C_s$ and $C_{s}^\dagger$
are Abelian, and thus, there are no systematic degeneracies in spectra
of all three Hamiltonians any more.  Correspondingly, the
doubly-degenerate eigenvalues of $\hat{H}_{-}^{\rm GP}$ and $\hat{H}$ 
split as $E_{1/2}= {B}_1 \oplus {B}_2$ \footnote{We use $B_1$ and $B_2$ representations
  of the $C_{2v}$ group to label the corresponding representations of
  the isomorphic double-group $C_{s}^\dagger$}.

{\it Eigenstate localization and symmetry breaking.---} Ground-state
degeneracy in the BO+GP and full models leads to the spontaneous
localization of the lowest eigenstates that has no analogs in the BO
model. It is well-known that the eigenstates of 
$\hat{H}_{-}^\text{BO}$ with the symmetric arrangement of wells
($\Delta = 0$) will be delocalized over the wells \cite{Takada:1994jo}. 
For high barriers, the low-lying eigenstates of $\hat{H}_{-}^\text{BO}$ 
have a group structure where states within a group are separated by a small energy gap, while
different groups are separated by large energy gaps. 
The ground and first excited states $\Psi^{\rm BO}_{1,2}$ 
of $H_{-}^{\rm BO}$ are delocalized functions corresponding to  
the lowest energy group. By rotating within the $\Psi^{\rm BO}_{1,2}$
subspace, one can obtain function $\Phi = \Psi^{\rm BO}_{1}\sin\eta+\Psi^{\rm BO}_{2}\cos\eta$ 
that is localized in the donor well. However, $\Phi$ is not an eigenfunction of $H_{-}^{\rm BO}$, and thus, 
it will escape from the donor well within a time period inversely proportional 
to the energy gap between the $\Psi^{\rm BO}_{1,2}$ eigenenergies.

To consider the full \emph{diabatic} problem [Eq.~(\ref{eq:diab-2st})] we introduce 
the lowest eigenstates $\Phi_{\rm D}$ and $\Phi_{\rm A}$ of the donor and
acceptor Hamiltonians $\hat{H}_{\rm D}=\hat{T}_N+V_{11}$ and $\hat{H}_{\rm A}=\hat{T}_N+V_{22}$, respectively.  
Within the full problem, vectors $(\Phi_{\rm D},~0 )^{\dagger}$ and $(0,~\Phi_{\rm A})^{\dagger}$ 
cannot interact via $V_{12}\sigma_x$ because both $\Phi_{\rm D}$
and $\Phi_{A}$ are even with respect to $y \to -y$, while
$V_{12}$ is odd. Therefore, true lowest eigenfunctions of the full problem
$\Psi_1^{\rm full}$ and $\Psi_2^{\rm full}$ are dominated by
the vectors $(\Phi_{\rm D},~0 )^{\dagger}$ and $(0,~\Phi_{\rm A})^{\dagger}$, while 
admixture of higher eigenfunctions of $\hat{H}_{\rm D}$ and $\hat{H}_{\rm A}$ 
is suppressed by an energy gap of at least $\omega_2$. Owing to the
degeneracy of the full problem spectrum, one can always rotate
$\Psi_{1,2}^{\rm full}$ into a pair of localized eigenfunctions
$\Psi_{\rm D, A}^{\rm full}$ that are close to the 
$(\Phi_{\rm D},~0 )^{\dagger}$ and $(0,~\Phi_{\rm A})^{\dagger}$ states. Similarly, rotating the lowest doublet
components $\Psi_{1,2}^{\rm GP}$ of $\hat{H}_{-}^{\rm GP}$ also produces
spatially localized eigenstates $\Psi_{D,A}^{\rm GP}$. Thus, 
the spectral degeneracy leads to localization of the lowest eigenstates and 
spontaneous symmetry breaking 
in both full and BO+GP models.

In the $\Delta \ne 0$ case, although the degeneracy of the spectra in the full and BO+GP models 
 is lifted, the localization of low-lying states survives in both models provided that $\Delta$ is not ``too large'' ($\Delta<\omega_i$). 
For the full model, $\Delta \ne 0$ is equivalent to introducing 
$\sigma_z\Delta/2$ perturbation to the $\Delta = 0$
Hamiltonian. This \emph{diagonal} perturbation does not couple the
localized eigenstates $\Psi_{\rm D,A}^\text{full}$ to the first order,
but only lifts the degeneracy by changing energies of these
states. Higher-order contributions from states other than $\Psi_{\rm
  D, A}^{\rm full}$ are energetically suppressed. 
In the GP+BO model, we start with a localized state of the donor 
well for the $\Delta=0$ problem $\Psi_{\rm D}^\text{GP}$ and
consider its dynamics in the $\Delta \ne 0$ case. Two minimal energy 
paths to the acceptor minimum are available for $\Psi_{\rm D}^\text{GP}$ via transition states TS$_1$ and
TS$_2$ (Fig.~\ref{fig:2d-sketch}b). Phases acquired by the wave packet
along these paths are close to $e^{i\pi/2} = i$ and $e^{-i\pi/2} =
-i$. Therefore, the interference between the parts of the wave packet that take different paths is
\emph{destructive} and leads to forming a nodal $y = 0$ line in the wave
packet~\cite{Schon:1994/cpl/55, Schon:1995bi}. A wave function with
an extra node has higher energy (approximately by $\omega_2$),
and thus, the $\Psi_{\rm D}^\text{GP}$ escape from the donor well is energetically 
suppressed when $\Delta<\omega_2$.

These qualitative considerations break down for sufficiently large
$\Delta$, therefore we perform numerical
simulations within the BO+GP approximation to find critical values of $\Delta$ when
localization disappears. For simplicity of the subsequent discussion only the isotropic case 
$\omega_1 = \omega_2=1$ will be considered, while its generalization 
for the $\omega_1 \ne \omega_2$ case is straightforward.  
Also, instead of an absolute value of the coupling $c$ we use the dimensionless parameter
$\gamma$. 

We separate donor and acceptor wells (Fig.~\ref{fig:2d-sketch}a)
using a projector operator ${\hat P}(x,y)$ that equals 1(0) if $(x,y)$
is in the donor(acceptor) well. For an eigenstate $\Psi$ 
the average value $P = \left\langle\Psi\right|{\hat P}\left|\Psi\right\rangle$ 
provides a quantitative measure of the $\Psi$ localization. 
In a doubly degenerate case $\Psi = \Psi_1\cos\eta + \Psi_2\sin\eta$, 
where $\Psi_{1,2}$ are orthogonal components of the eigen-subspace, 
we define the $\Psi$ localization as $\max_{\eta}P$. 

\begin{figure}
  \centering
  \includegraphics[width=0.50\textwidth]{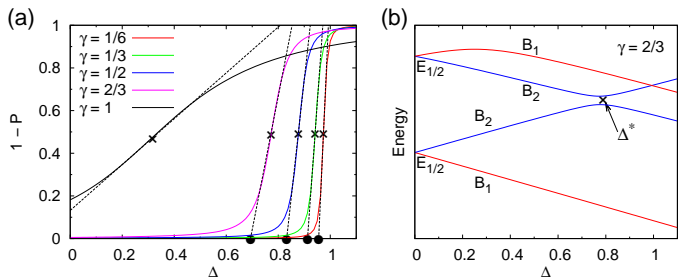}
  \caption{a) Delocalization degree ($1-P$) for the $\hat{H}_{-}^{\rm GP}$ first excited state as a function of $\Delta$.  Crosses mark
    inflection points, dashed lines are tangent lines at  the inflection points. Filled circles separate the regions of slow and
    fast growth of $P$. b) Correlation diagram for low-energy eigenstates of $\hat{H}_{-}^{\rm GP}$. The cross marks the
    energy gap minimum of two states in the avoided-crossing and corresponds to the inflection point of the
    $1-P$ ($\gamma = 2/3$) curve on panel (a).}
  \label{fig:b2acc_curv} 
\end{figure}

Figure~\ref{fig:b2acc_curv}a presents the delocalization degree $1-P$ 
of the first excited state as a function of $\Delta$~\footnote{The ground state is
  trivially localized for $\Delta \ne 0$ since we always assume that
  the acceptor is \emph{lower} in energy than the donor.}.
The delocalization quickly raises to 1 when $\Delta$ (for a given
$\gamma$) passes through a certain critical value
$\Delta^{*}(\gamma)$. This behavior can be explained by considering
the eigenvalue correlation diagram (Fig.~\ref{fig:b2acc_curv}b): at
$\Delta^{*}$ the first excited state exhibits an avoided-crossing with
the second excited state of the same symmetry but localized in
the acceptor well. The donor state restores its localization beyond the
critical point when it becomes the second excited state. However,
after that point its energy grows beyond the region relevant to low-energy
dynamics.

\begin{figure}
  \centering
  \includegraphics[width=0.50\textwidth]{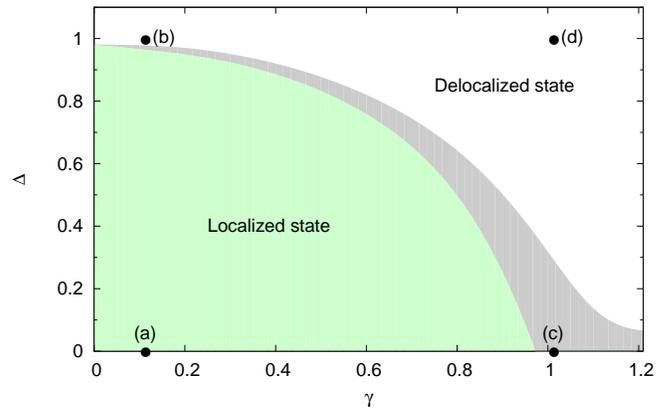}
  \caption{Phase diagram based on the critical values of $\Delta$ from Fig.~\ref{fig:b2acc_curv}a:
the inflection points of the $1-P$ curves (grey-white boundary), the intersections of the tangent lines with the $\Delta$ axis (green-grey boundary). 
Black dots (a-d) mark the parameter values for which nuclear dynamics is presented in Fig.~\ref{fig:xi-eq-0}(a-d).} 
  \label{fig:phase_diag}
\end{figure}

Based on the shape of $1-P$ curves (Fig.~\ref{fig:b2acc_curv}a), we
propose two definitions of $\Delta^{*}$: 1) the inflection point of
the $1-P$ curve, and 2) the intersection of the tangent line at the
inflection point with the $\Delta$ axis. Since we consider finite systems
where true continuous phase transitions between
localized and delocalized states are impossible, these two definitions of $\Delta^{*}$
give different estimates for transition points. We put both critical values
of $\Delta^{*}$ for different $\gamma$ on the same plot and obtain the
phase diagram for $\hat{H}_{-}^\text{GP}$ in Fig.~\ref{fig:phase_diag}.
Colored areas correspond to a range of couplings ($\gamma$) and
acceptor shifts ($\Delta$) for which the donor---despite unfavorable
energetics---still supports the localized eigenstate. 

\begin{figure}
  \centering
  \includegraphics[width=0.50\textwidth]{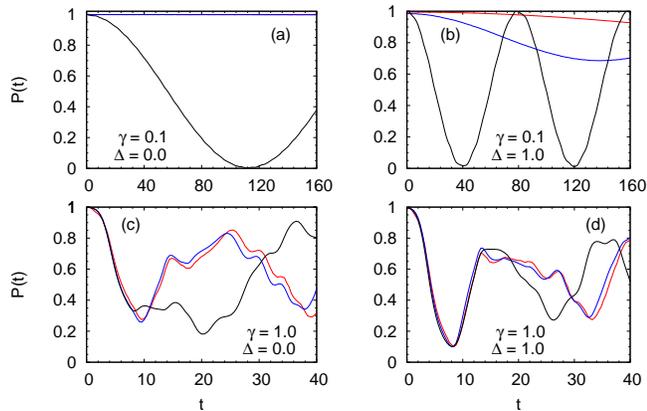}
  \caption{Population transfer dynamics for $T = 0$: the full model (red), the BO+GP model (blue),
    the BO model (black).}
  \label{fig:xi-eq-0}
\end{figure}

{\it Nuclear dynamics.---} To elucidate the impact of GP and
localization on the dynamics, we simulate the probability transfer
$P(t) = \tr\left\{{\hat \rho}(t) {\hat P}\right\}$ starting from the
Boltzmann density of the donor Hamiltonian $\hat{H}_{\rm D}$ at
temperature $T$ (simulation details are given in Ref.~\onlinecite{Note4}). 
The localization of eigenstates in the presence of GP has a large impact on the nuclear dynamics. 
At $T=0$, in the presence of the localization in $\hat{H}^{\rm GP}_{-}$, our initial state can be close 
to stationary by almost coinciding with a single eigenstate of $\hat{H}^{\rm GP}_{-}$. 
In contrast, in the BO model where the localization is impossible, 
the initial wave packet is predominantly a superposition of two lowest BO eigenstates. 
Indeed, at $\gamma = 0.1$ and $\Delta = 0$ [point (a) in Fig.~\ref{fig:phase_diag}] both full and BO+GP models 
demonstrate complete suppression of the tunnelling, while the BO model 
produces unit size coherent oscillations according to the Rabi two-level model (Fig.~\ref{fig:xi-eq-0}a).
When we are slightly above the edge of
stability of the localized phase in the BO+GP problem 
[$\gamma = 0.1$ and $\Delta = 1$, point (b) in  Fig.~\ref{fig:phase_diag}], 
the dynamics of all three models are significantly different, 
and the BO+GP dynamics is faster than those of the full model (Fig.~\ref{fig:xi-eq-0}b). 
The analysis of the eigenvalue correlation diagram for the full model reveals that the
corresponding critical $\Delta^{*}$ for $\gamma=0.1$ is slightly
larger than that of the BO+GP model. Thus, the dynamics represents only the onset of
the localized state decay in the full model, while for the BO+GP model
the localized state is already unstable. Except for a rather narrow
range of $\Delta$'s the full and BO+GP models are quite similar (Fig.~\ref{fig:xi-eq-0}a,c,d). The
role of non-adiabatic transitions is indeed small,
and all observed effects can be attributed to the presence of GP. 
Finally, in the delocalized region [point (d) in Fig.~\ref{fig:phase_diag}] 
where the initial localized state is unrepresentable as a single eigenstate, 
all three models demonstrate almost quantitatively 
similar dynamics  with quick and profound population transfer (Fig.~\ref{fig:xi-eq-0}d).  

\begin{figure}
  \centering
  \includegraphics[width=0.5\textwidth]{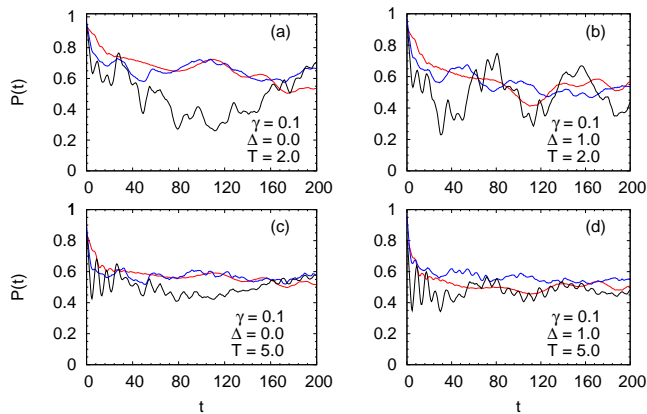}
  \caption{Population transfer dynamics for finite $T$'s: the full model (red), the BO+GP model (blue),
    the BO model (black).}
  \label{fig:T-ne-0}
\end{figure}

Our non-zero $T$ set-up corresponds to quick thermalization of the
initial state by environment followed by system dynamics that does not 
account for interaction with environment. Temperature
averaging includes several states that may have different degrees of
localization. If we populate levels with a similar localization
pattern, $T \ne 0$ dynamics preserves the qualitative features of that
for $T=0$ ({\it cf.}  Fig.~\ref{fig:xi-eq-0}a-b with Fig.~\ref{fig:T-ne-0}a-b), whereas differences in the
localization character of populated levels create differences in
dynamics ({\it cf.} Fig.~\ref{fig:xi-eq-0}a with Fig.~\ref{fig:T-ne-0}a).  Increasing temperature even more will
eventually populate enough levels to wash out all state-specific
dynamical features (Fig.~\ref{fig:T-ne-0}c-d).

{\it Extensions of the model.---} Understanding dynamics with GP for
more complex Hamiltonians can be accomplished via perturbative
consideration of the corresponding two-state problem. Two main
factors are: the selection rules for the coupling matrix elements between
 diabatic vibrational levels, and energy differences between coupled levels. 
For more general coupling potentials $V_{12} \sim y^n\sigma_x$ odd $n$'s are expected to produce
similar effects to those of $n=1$ as they can support the spinor symmetry,
while even $n$'s cannot and their behavior is closer to the case of a
constant coupling potential. Other ways to modify selection rules also include
introducing different frequencies for the donor and acceptor,
non-orthogonal tuning and coupling coordinates, Dushinsky rotation or
anharmonicity of wells. All these modifications unless they
are too strong can be analyzed perturbatively.

In conclusion, we investigated dynamical consequences of the spontaneous symmetry breaking 
in the two-dimensional CI model. The naive BO approximation 
cannot capture the correct spinor symmetry of the problem and 
breaks down qualitatively even for the nuclear dynamics in regions that are far from the CI. 
Introducing GP explicitly into the nuclear wavefunction restores the symmetry and 
the associated symmetry breaking. The latter leads to the spatial localization of vibronic eigenstates
and freezes the inter-well dynamics. Owing to the topological character of this effect one can see this 
as an example of a topological insulating state \cite{Hasan:2010ku} in a finite system. 
Variation of parameters can lower the overall symmetry and remove the  
symmetry breaking; however, the localization of the eigenstates and its dynamical consequences 
persist in some range.
 \begin{acknowledgements}
We thank L. A. Pachon, T. V. Tscherbul, S. Wittington,  V. V. Albert, J. Endicott, and P. Brumer 
for helpful discussions and comments on the manuscript.  
This work was supported by NSERC of Canada through the Discovery Grants Program.
\end{acknowledgements}

\end{document}